\documentclass[twocolumn,english,aps,prb,showpacs,byrevtex,amsmath,amssymb,superscriptaddress,shortbibliography]{revtex4-2}

\usepackage{graphicx}
\usepackage{epstopdf}
\usepackage{dcolumn}
\usepackage{bm}
\usepackage{gensymb}
\usepackage{upgreek}
\usepackage{hyperref}
\usepackage{graphicx}
\usepackage{dcolumn}
\usepackage{bm}
\usepackage{gensymb}
\usepackage{upgreek}
\usepackage{booktabs}
\usepackage{hyperref}

\usepackage{amssymb,mathtools}
\usepackage{color}
\usepackage{amsmath} 

\usepackage{tikz,xcolor,hyperref}

\definecolor{lime}{HTML}{A6CE39}
\DeclareRobustCommand{\orcidicon}{%
	\begin{tikzpicture}
	\draw[lime, fill=lime] (0,0)
	circle [radius=0.16]
	node[white] {{\fontfamily{qag}\selectfont \tiny ID}};
	\draw[white, fill=white] (-0.0625,0.095)
	circle [radius=0.007];
	\end{tikzpicture}
	\hspace{-2mm}
}

\foreach \x in {A, ..., Z}{%
	\expandafter\xdef\csname orcid\x\endcsname{\noexpand\href{https://orcid.org/\csname orcidauthor\x\endcsname}{\noexpand\orcidicon}}
}

\begin{document}

\title{Dirac edge states of two-dimensional altermagnetic topological crystalline insulators}

\author{Raghottam M. Sattigeri\orcidC}
\email{raghottam.sattigeri@polimi.it}
\affiliation{Physics Department, Universit\'a degli Studi di Milano, Via Celoria 16, 20133 Milan, Italy}

\author{Xujia Gong\orcidE}
\affiliation{International Research Centre Magtop, Institute of Physics, Polish Academy of Sciences,
Aleja Lotnik\'ow 32/46, 02668 Warsaw, Poland}

\author{Amar Fakhredine\orcidF}
\affiliation{Institute of Physics, Polish Academy of Sciences, Aleja Lotnik\'ow 32/46, 02668 Warsaw, Poland}
\affiliation{International Research Centre Magtop, Institute of Physics, Polish Academy of Sciences, Aleja Lotnik\'ow 32/46, 02668 Warsaw, Poland}

\author{Carmine Autieri\orcidA}
\affiliation{International Research Centre Magtop, Institute of Physics, Polish Academy of Sciences, Aleja Lotnik\'ow 32/46, 02668 Warsaw, Poland}
\affiliation{SPIN-CNR, UOS Salerno, IT-84084 Fisciano (SA), Italy}

\author{Giuseppe Cuono\orcidB}
\affiliation{Consiglio Nazionale delle Ricerche (CNR-SPIN), Unità di Ricerca presso Terzi c/o Università “G. D’Annunzio”, 66100, Chieti, Italy}

\begin{abstract}
Two-dimensional (2D) metallic altermagnets present exciting opportunities for both fundamental research and practical innovations. Their ability to enhance tunneling magnetoresistance in magnetic tunnel junctions, combined with the direct control of spin currents via electric fields, makes them highly promising for spintronic devices. Moreover, the rich electronic structure of altermagnets can host nontrivial topological phases. 
In particular, topological crystalline insulators are compounds where the topological states are protected by both crystalline and time-reversal symmetries. Furthermore, manipulating the state of a system between topological and trivial phases through external parameters unlocks new possibilities for quantum materials and advanced electronics.
We show the edge states of a 2D altermagnetic topological crystalline insulator, using as a representative example Cr$_2$BAl, a  2D MBene metallic altermagnet with a d$_{x^2-y^2}$ altermagnetic ordering.
We find that the system  can host an altermagnetic phase with extremely large ``weak ferrimagnetism" which is sizeable also with light atoms, only with an in-plane component
of the N\'eel vector. The electronic structure of Cr$_2$BAl presents multiple crossings and anti-crossings in the
vicinity of the Fermi level along [100] and [010] directions. When the spin-orbit coupling interaction is included, with the N\'eel vector along [001] direction, energy gaps open at the band crossing points, resulting in a pronounced peak in the spin Hall conductivity.
The simulated Cr-B terminated [100] edge-projected band structure reveals Dirac dispersions at the bulk crossings and anti-crossings, which are absent in Cr-Al terminations. 
\end{abstract}

\maketitle


Magnetic materials are widely studied for their applications. Ferromagnets exhibit net magnetization and spin-split bands, while antiferromagnets show zero net magnetization and spin-degenerate bands. Recently discovered altermagnets combine these properties, featuring spin-split bands in momentum space and zero net magnetization in real space \cite{Smejkal22,Smejkal22b,mazin2022altermagnetism}. This spin-splitting arises without spin-orbit coupling (SOC), making it non-relativistic. Symmetry-wise, antiferromagnets rely on inversion or translation symmetry, whereas altermagnets rely on rotational or mirror symmetries. Including SOC, altermagnets exhibit an anomalous Hall effect (AHE) even without ferromagnetic order \cite{Smejkal20,Shao20}, with weak ferromagnetism dependent on the magnetic moment direction \cite{Shao20,autieri2023dzyaloshinskii}. In systems with staggered Dzyaloshinskii-Moriya interaction, spin components orthogonal to the N\'eel vector can induce weak ferromagnetism or weak ferrimagnetism \cite{autieri2023dzyaloshinskii}. Thanks to these properties, altermagnets hold promise for spintronics and quantum information technologies \cite{Shao21,Hernandez21,GiantMagneto22,zhou2023crystal,Ouassou23}.

Altermagnetism has been experimentally observed in bulk materials \cite{Krempasky24,Zhu24} and it has been predicted in many 3D compounds \cite{Smejkal22,Smejkal22b,Smejkal23,Mazin21,Mazin23,Cuono23JMM,Fakhredine23,Grzybowski24}.
Achieving altermagnetism in 2D materials is more challenging compared to 3D compounds. This is primarily because 2D systems are confined to a plane, allowing only in-plane (planar) spin–momentum locking. In contrast, 3D materials offer additional momentum components and allow for bulk spin–momentum locking. Furthermore, when we project the 3D BZ in a 2D one, the $k$-points with opposite spin-splitting can merge and cancel the altermagnetism \cite{Sattigeri23}.
Research is ongoing on 2D altermagnets, which can be very useful for many applications. Altermagnetism was predicted in RuF$_4$ \cite{Milivojevic24}, FeS(110) \cite{Liu24} and in ultrathin films 
of GdAlSi \cite{Parfenov25}, it was shown that Fe$_2$Se$_2$O and V$_2$SeTeO can present piezo-altermagnetism \cite{Zhu24piezo,Wu24}, while it was experimentally realized in V$_2$Se$_2$O \cite{Ma21} and KV$_2$Se$_2$O \cite{Jiang24}.
2D altermagnets can be used as electrodes in magnetic tunnel junctions (MTJ) to increase the tunneling magnetoresistance (TMR) effect \cite{Shao21}.
Furthermore, it was demonstrated that some 2D antiferromagnets can be functionalized into altermagnets by using an external electric field \cite{Wang25,Mazin23monolayer} and it was shown that the breaking of time-reversal symmetry in the Fe-based superconducting chalchogenides can be due to altermagnetism \cite{Mazin23monolayer}, highlighting the potential to explore the interplay between 2D altermagnetism, superconductivity and topology.
It was shown that Cr-based 2D metal borides, namely MBenes layers, can be used in the design of 2D altermagnets promoting intriguing phenomena \cite{Gu21, Sun25}. 
\\

Topological insulators are a distinct phase of matter with an insulating bulk and conductive surface or edge states protected by time-reversal symmetry \cite{Hasan10}. These states feature a Dirac cone, mimicking the energy-momentum relation of relativistic particles.
Topological crystalline insulators (TCIs) exhibit surface states that are protected by both time-reversal symmetry and crystalline symmetries \cite{Fu11,Hsieh12}. Unlike topological insulators, which host a Dirac point at $\overline{\Gamma}$, TCIs exhibit it at other high-symmetry points of the surface BZ, depending on the crystal symmetry. The discovery of the TCI phase in SnTe and its alloys has fueled extensive research into their properties \cite{Dziawa12}.
The mirror Chern number is a topological invariant that characterizes systems with mirror symmetry, ensuring the presence of protected edge states in topological crystalline insulators \cite{Hsieh12}.
\\

Microscopic models for 2D d$_{xy}$ or d$_{x^2-y^2}$ altermagnets have demonstrated the appearance of previously unobserved Dirac crossings protected by mirror symmetries between bands with the same spin orientation but located on opposite sub-lattices. When the Néel vector is oriented perpendicular to the plane, the spin-orbit coupling induces a gap at these Dirac points, giving rise to mirror Chern bands which enables the quantum spin Hall effect \cite{antonenko2024mirrorchernbandsweyl}.  The symmetry conditions enabling topologically protected Dirac points in 2D altermagnets were studied \cite{Parshukov25}.
Other topological properties were discussed in 3D altermagnets \cite{Ma24}, including  topological features in the antiferromagnetic metallic ground state of the FeSb$_3$ skutterudite \cite{dilucente2025spinfluctuationssteerelectronic}, Dirac fermions in the altermagnet Ce$_4$Sb$_3$ \cite{He25}, and the discovery of Weyl altermagnetism \cite{Li24,Lu25,Nag24,Qu25} in CrSb and GdAlSi using experimental methods like spin angle resolved photoemission spectroscopy or ab-initio simulations and models.
\\

\begin{figure}
	\includegraphics[width=0.5\textwidth]{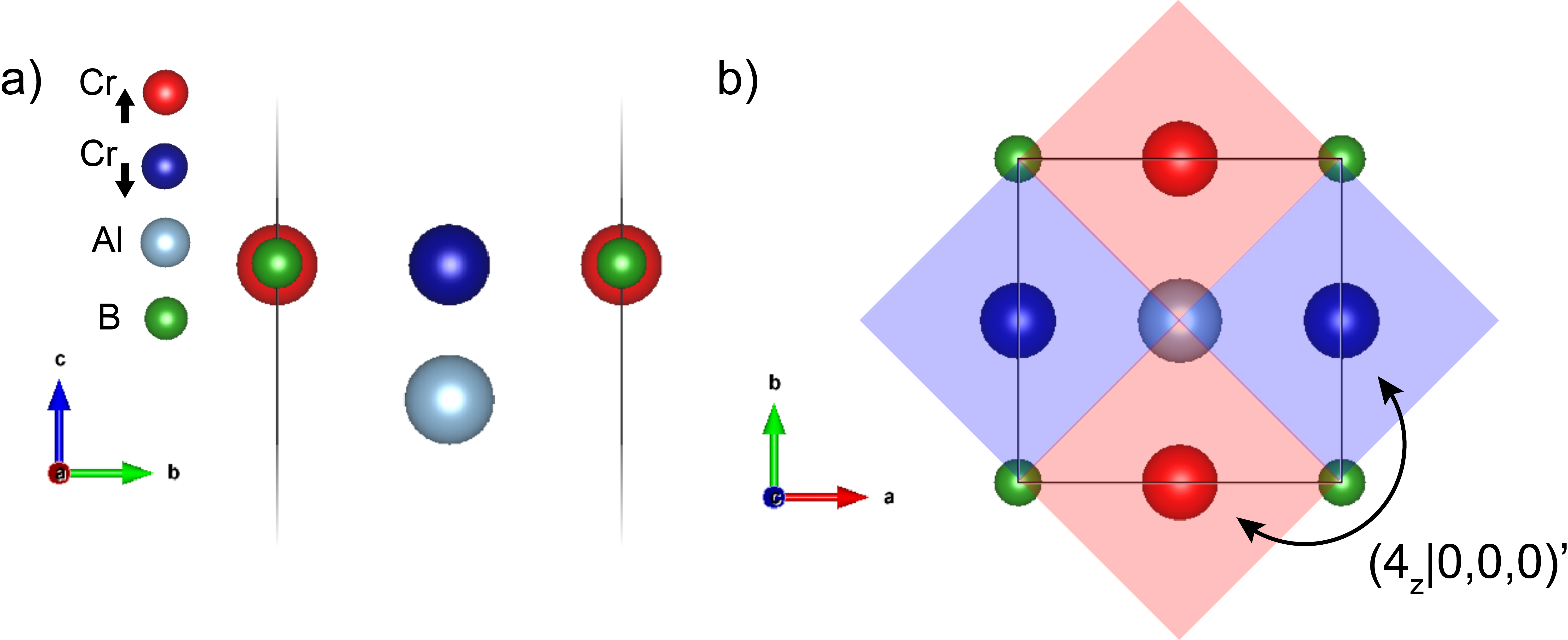}
	\caption{The side view \textbf{(a)} and the top view \textbf{(b)} of the 2D Cr$_2$BAl unit cell. The chromium sub-lattice with opposite spin polarization are distinguished by blue and red balls, boron atoms in green and aluminum in gray. The rectangle and arrow in blue and purple highlight the generator of opposite-spin sub-lattice rotations. Here, the rotational operation was labeled by Litvin's notation (4${_z}|$0,0,0)'.}
	\label{fig:figure1}
\end{figure}

\begin{figure}[ht!]
	\centering
	\includegraphics[width=0.45\textwidth]{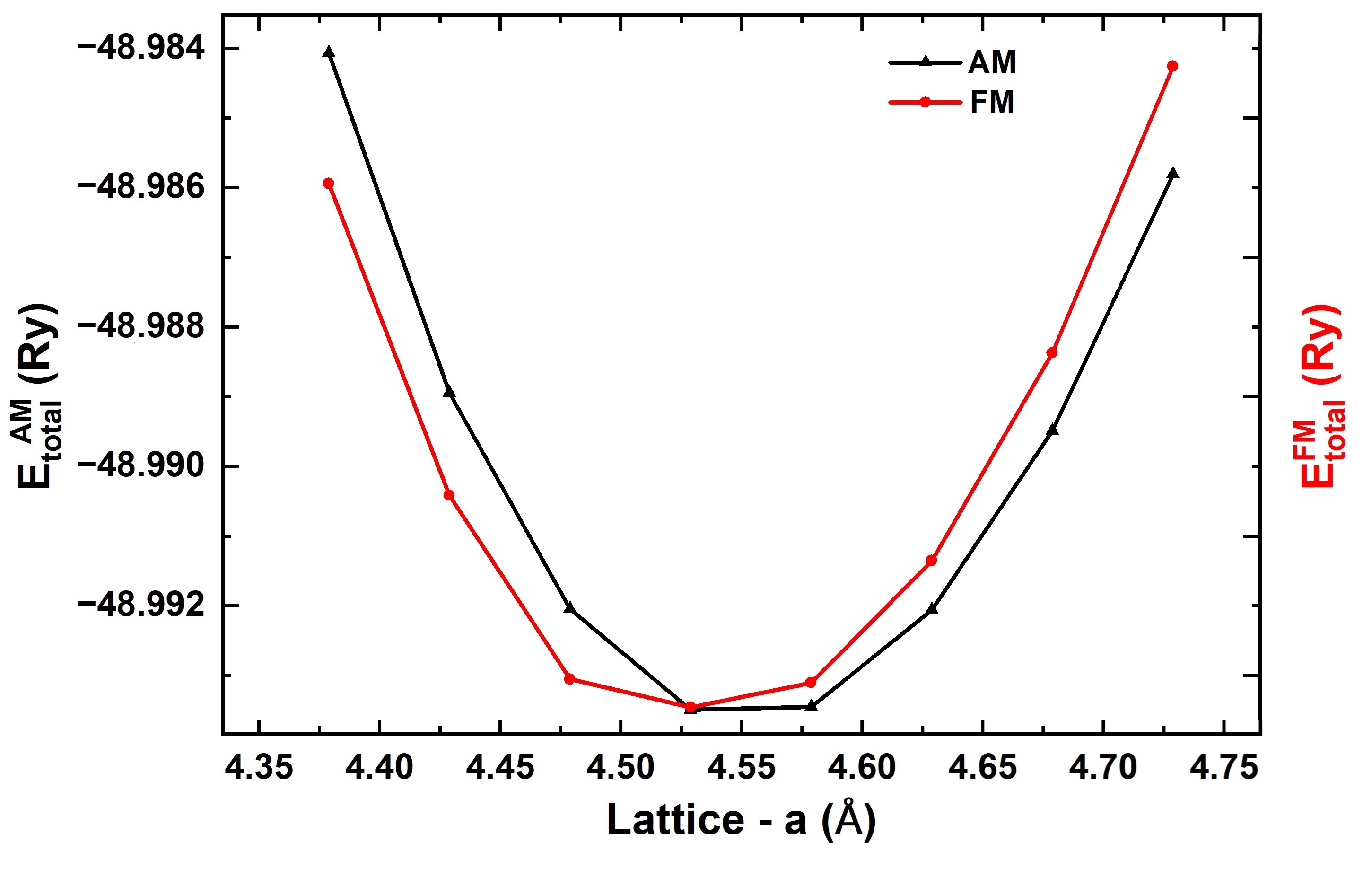}
	\caption{The total energy for both the AM and FM configurations as a function of the lattice parameter $a$. The AM configuration exhibits a lower energy at around 4.54 \AA, indicating a more stable magnetic ground state. A biaxial strain in the a-b plane can induce an altermagnet-ferromagnet transition.}
	\label{fig:figure2}
\end{figure}

In this work, we aim to predict the properties of a new material, Cr$_2$BAl, which belongs to the altermagnetic family. Some members of this family were recently synthesized \cite{https://doi.org/10.1002/admi.202500092}. 
We employ Density Functional Theory (DFT) calculations to determine the electronic structure and magnetic properties of the 2D monolayer Cr$_2$BAl. It is a P-2 \cite{Smejkal22b} and a $d$-wave altermagnet, with Dirac edge states away from the $\bar{\Gamma}$ point, therefore, an altermagnetic topological crystalline insulator. To our knowledge, this is the first work where the edge states of an altermagnetic topological crystalline insulator are shown, and, while other works use microscopical models to analyze theoretically the topological properties, we investigate, by means of computational techniques, a material that can be experimentally realized.
In the following, we present the computational details and the structural, electronic and topological properties, concluding with the final remarks. 
\\

\begin{figure*}[ht!]
	\centering
	\includegraphics[width=\textwidth]{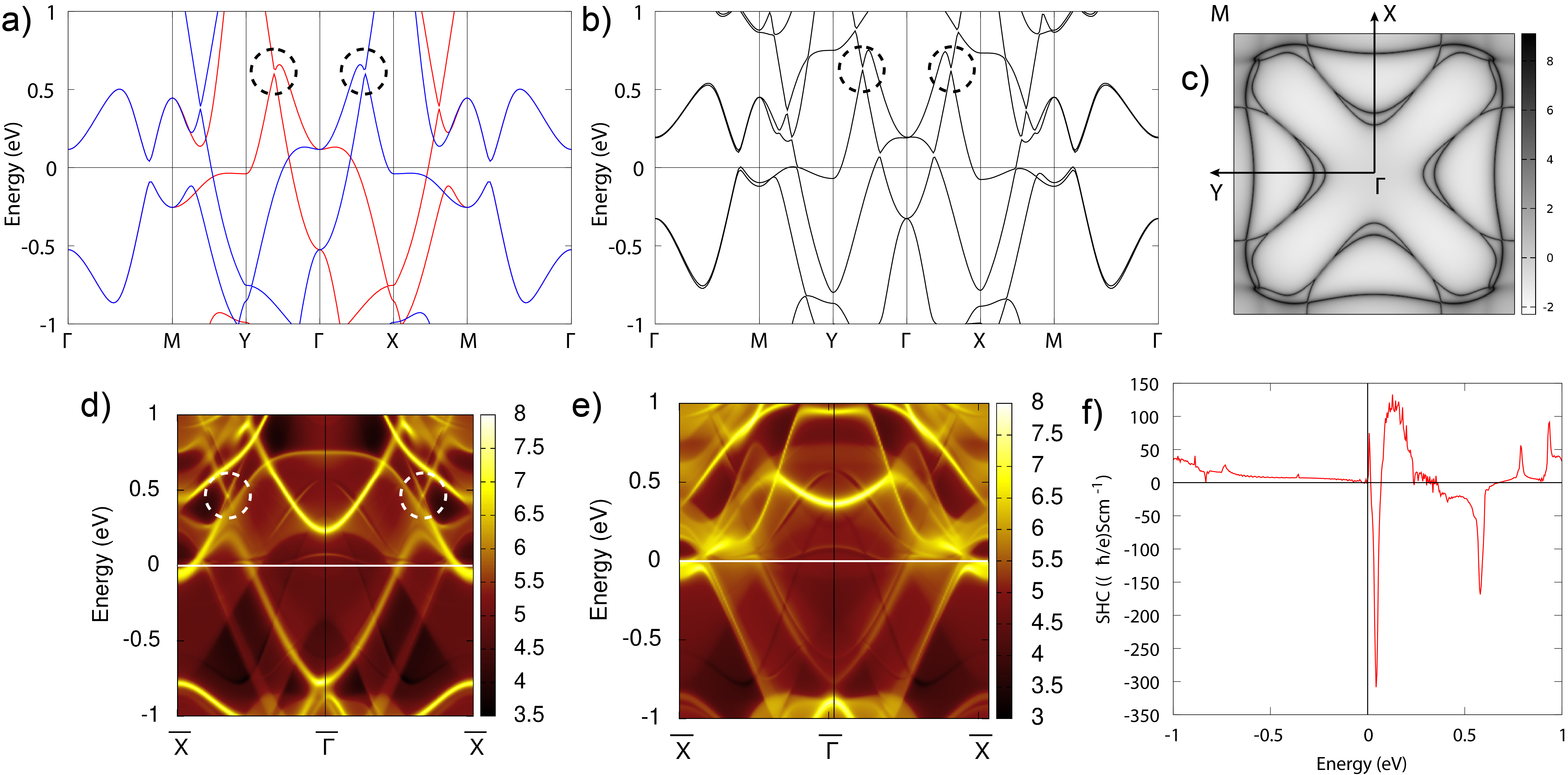}
	\caption{\textbf{(a)} The non-relativistic spin-splitting (red is for spin-up and blue is for spin-down) in electronic band structure for Cr$_2$BAl monolayer with band crossings and anti-crossings (highlighted by dashed circles) along the BZ path X-$\Gamma$-Y, i.e., in [100] and [010] directions. \textbf{(b)} The electronic structure obtained with spin-orbit coupling (with N\'eel vector in [001] direction) indicates a spin-orbit interaction-induced gap (highlighted by dashed circles) along the BZ path X-$\Gamma$-Y, i.e., in [100] and [010] directions. \textbf{(c)} Fermi surface slice in the 2D BZ, corresponding to the spin-orbit coupling bands presented in \textbf{(b)}. Surface-projected band structure obtained using surface projections in [100] direction: \textbf{(d)} The Cr-B surface termination indicates the presence of Dirac dispersions (highlighted by dashed circles) which coincide with the relativistic spin-splittings observed in \textbf{(b)}, leading to the peak in spin Hall conductivity at $\sim$ 525 meV observed in \textbf{(f)}. \textbf{(e)} The Cr-Al surface termination, on the contrary, indicates the absence of the Dirac dispersions. \textbf{(f)} The spin Hall conductivity indicating sharp peaks at $\sim$ 525 meV and $\sim$ 100 meV, corresponding to the multiple band crossings/anti-crossings in the non-relativistic regime and the spin-orbit induced splitting in the relativistic regime presented \textbf{(a)} and \textbf{(b)}, respectively. The Fermi level is set to zero in all plots.}
	\label{fig:figure3}
\end{figure*}



DFT was employed to compute the electronic structure, edge states and spin Hall conductivity as implemented in Quantum ESPRESSO with post-processing using Wannier90 and WannierTools \cite{Giannozzi17,Marzari12,Mostofi14,Wu18,sancho1985highly}.
We used the scalar-relativistic norm-conserving Martins-Troullier pseudopotentials with generalized-gradient approximations and Perdew-Burke-Ernzerhof exchange-correlation functional with optimized kinetic energy cut-off of 80 Ry \cite{Blochl94,Perdew96}. Full relativistic nonlinear core corrections were included in the pseudopotential to address spin-orbit coupling effects. A uniform Monkhorst-Pack grid of 10 $\times$ 10 $\times$ 1 was used to represent $k$-points in the irreducible two-dimensional Brillouin zone (BZ) in all the calculations \cite{Monkhorst76}. Van-der-Waals Grimme corrections were included in all the calculations to have accurate energetic description of the system \cite{grimme2006semiempirical}.
To accurately describe the $d$-orbital electrons of transition metals, the DFT+U Dudarev formalism has been employed, with the Hubbard U term to simulate the electron correlation effects by using U = 4 eV and ortho-atomic projections.
This was followed by wannierization and disentanglement procedure, which involved the $d$-orbitals localized on Cr sites and the $p$-orbitals localized on B and Al sites using Wannier90. For the $d$-orbitals, we have considered only the majority electrons for each atom. The edge states and spin Hall conductivity were computed using the iterative Green’s function method implemented in WannierTools.
\\



The side and top views of the 2D Cr$_2$BAl monolayer are shown in Figures \ref{fig:figure1}(a) and \ref{fig:figure1}(b). The Cr$_2$BAl unit cell comprises two chromium atoms, one boron atom, and one aluminum atom, forming a tetragonal magnetic space group (MSG) P4$'$m$'$m (BNS$\#$99.165, OG$\#$99.3.825). This structure consists of two layers: one composed of aluminum atoms and the other composed of chromium and boron atoms, which have a minimal difference in their $z$-coordinates.
The two chromium atoms within the unit cell belong to different sub-lattices, denoted as Cr$_\uparrow$ (red) and Cr$_\downarrow$ (blue) as shown in Fig. \ref{fig:figure1}(b). Each chromium atom exhibits a magnetic moment of 4.095 $\mu_B$ along the easy axis in the (001) direction (out-of-plane), with a magnetic anisotropy energy (MAE) of 81 $\mu$eV per f.u. This is relatively high due to the anisotropy of the 2D system, indeed, the MAE is an order of magnitude higher than that of some conventional bulk magnetic materials like Fe (1.4 $\mu$eV per atom) and Ni (2.7 $\mu$eV per atom), indicating the magnetic moment is stably maintained along the easy axis and allowing the long-range magnetic order to persist at finite temperature \cite{Daalderop90,Fang18}.
We performed a scan of the energies of the altermagnetic (AM) and ferromagnetic (FM) configurations as a function of the lattice constant $a$, as reported in Fig. \ref{fig:figure2}. There is a broad range of values of $a$ where the AM configuration is the ground state.



Unlike conventional antiferromagnets, where sub-lattices with opposite spin polarization are coupled by translation and/or inversion symmetries, the sub-lattices in this system are coupled by mirror and rotational symmetries. This results in an electronic structure that breaks time-reversal ($\mathcal{T}$) symmetry, leading to alternating spin polarization in both coordinate and momentum spaces \cite{Litvin01,Hayami19}.
The electronic band structure is shown in Fig. \ref{fig:figure3}(a), exhibiting a typical altermagnetic pattern given that the $\mathcal{T}$ symmetry is broken.  The electronic bands are highly spin-splitted along $\Gamma$ - X and $\Gamma$ -Y, which is expected to generate large TMR, while they are degenerate along $\Gamma$-M.
The system is metallic and it presents a $d_{x^2-y^2}$-wave altermagnetism, with opposite spin-splitting along $\Gamma$ - X and $\Gamma$ - Y.
We show in Fig. \ref{fig:figure3}(b,c) the bands with SOC and Fermi surfaces, where it can be seen that they are linked by rotational symmetry.


\begin{figure*}
\centering
\includegraphics[width=0.227\linewidth,angle=270]{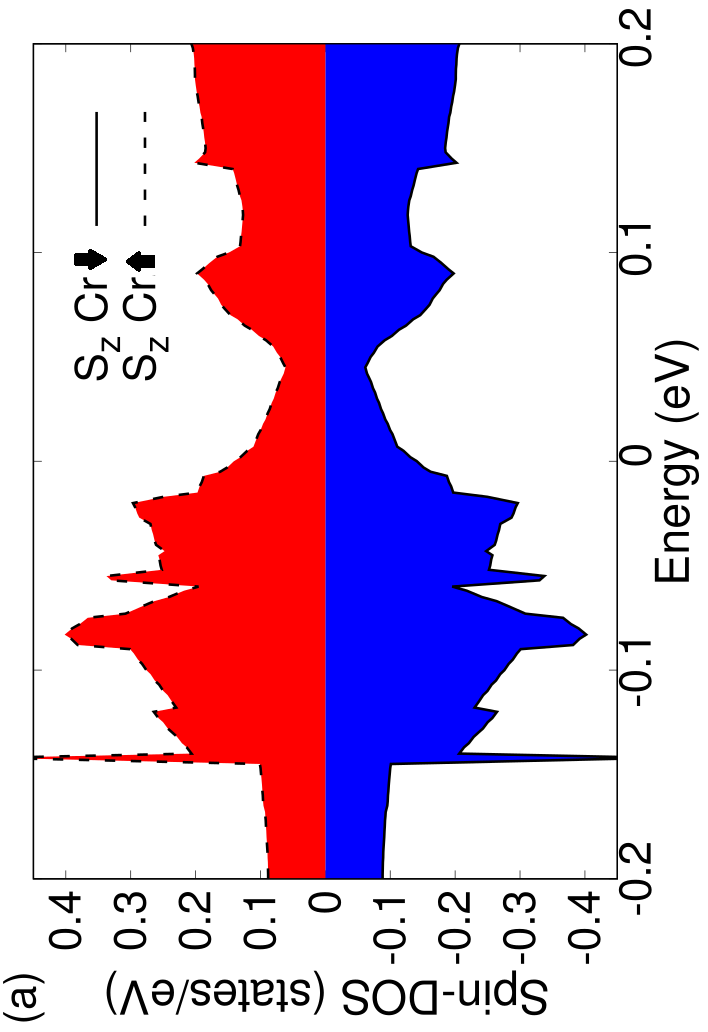}
\includegraphics[width=0.227\linewidth,angle=270]{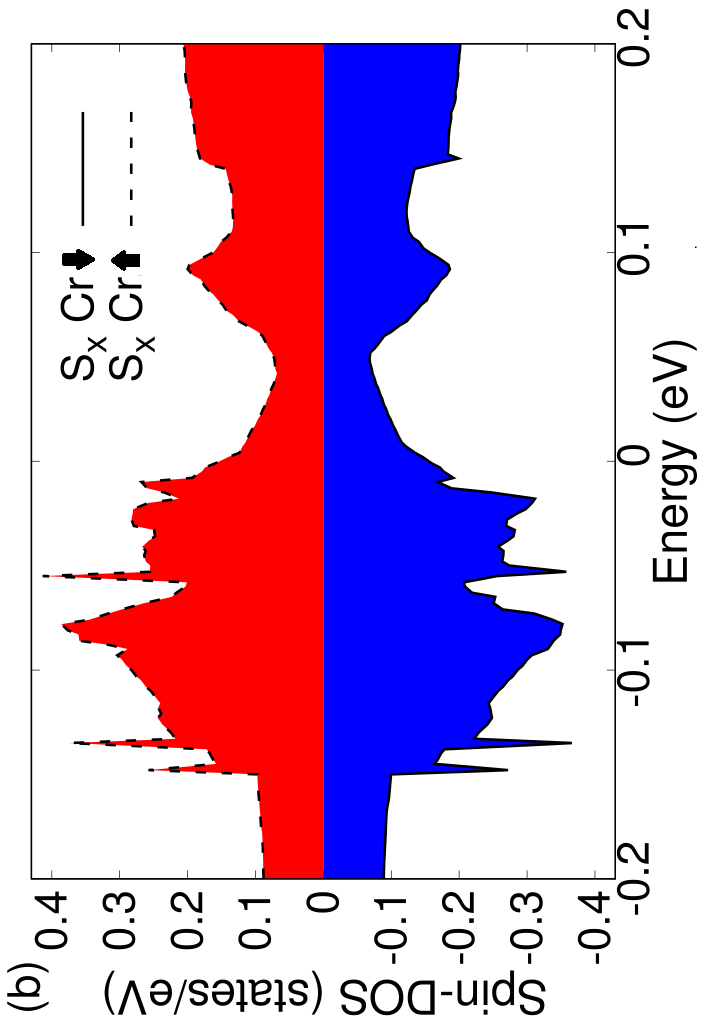}
\includegraphics[width=0.227\linewidth,angle=270]{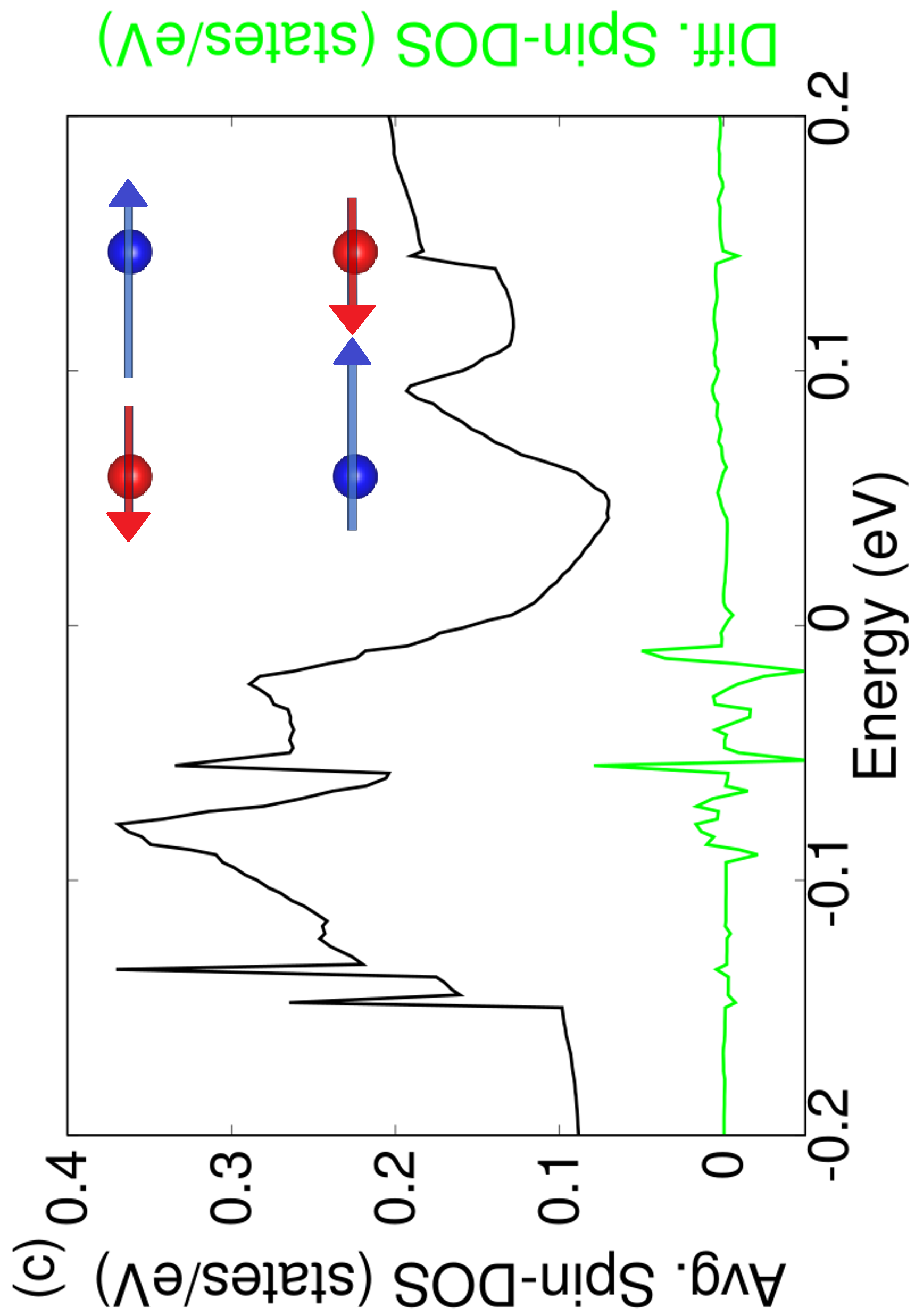}
\caption{Spin density of states of Cr$_\uparrow$ and Cr$_\downarrow$ at the low-energy for Cr$_2$BAl with N\'eel vector (a) along the $z$-axis and (b) along the $x$-axis. (c) Average spin-density and difference in the spin-density for the N\'eel vector along the $x$-axis. In the inset of panel (c), we report the magnetic spin configuration; the difference between the spins is exaggerated for graphical purposes. The difference between the spin densities of Cr$_\uparrow$ and Cr$_\downarrow$ represents the hallmark of weak ferrimagnetism.}
\label{fig:spindensity}
\end{figure*}

Although the SOC strength of Cr is not as high as in conventional heavy elements, it is not negligible either. We find that, in the non-relativistic regime, the altermagnetic bands (see Fig. \ref{fig:figure3}(a)) possess multiple crossings and anti-crossings in the vicinity of the Fermi level along [100] and [010] directions. This mandates a follow-up investigation of the influence of spin-orbital interactions on such crossings and anti-crossings, leading to insights on spin transport phenomena such as spin Hall conductivity.
With this objective in mind, we performed band structure calculations in the presence of SOC interactions. Since it was shown that the behavior of these crossings when SOC is introduced depends on the direction of the N\'eel vector and the mirror Chern bands appear for an out-of-plane N\'eel vector, we focus on the case when its direction is [001] \cite{antonenko2024mirrorchernbandsweyl}. The electronic band structure in the presence of SOC and with the N\'eel vector along [001] direction is presented in Fig. \ref{fig:figure3}(b). It is evident that, due to SOC, band gaps are opened in correspondence with the crossings and anticrossings of the bands, and they are of the order of $\sim$ 10-26 meV. This implies that we can exploit this feature in low-temperature experimental scenarios by varying the carrier concentrations and tuning the chemical function. This splitting is known to generate and host Dirac dispersions. We observe a clear signature of Dirac dispersions from ARPES-like edge states projected onto [100] surface and presented in Fig. \ref{fig:figure3}(d). However, it is imperative to note that surface terminations are vital in such cases. We can see that the Dirac dispersions persist only for Cr-B terminations, whereas they are absent for Cr-Al terminations (see Fig. \ref{fig:figure3}(e)).
The system is a topological crystalline insulators \cite{Hsieh12,Dziawa12}, and
the small relativistic spin-splitting manifests itself as a peak in the spin Hall conductivity plot, which is calculated as a function of the energy and presented in Fig. \ref{fig:figure3}(f).
\\



The non-relativistic spin-splitting phenomena observed in altermagnetic materials are governed by the direction of the N\'eel vector, especially in materials where SOC interactions are non-negligible. The spin canting effects originating from Dzyaloshinskii–Moriya interactions need to be addressed since they give rise to weak ferromagnetism or ferrimagnetism, depending on the presence or absence of centrosymmetry in the material, respectively \cite{autieri2023dzyaloshinskii}.
Cr$_2$BAl is governed by the tetragonal symmetry, which implies an absence of centrosymmetry.
Using the same technique on spin-density of state developed by us to study the spin-orbital effect in altermagnets \cite{autieri2023dzyaloshinskii}, we demonstrate the presence of weak ferrimagnetism in this system only with the in-plane component of the N\'eel vector. First, we calculate the spin-density for the spin along the $z$-direction represented in Fig. \ref{fig:spindensity}(a). The spin-densities of S$_z$ for Cr$_\uparrow$ and Cr$_\downarrow$ are equal and opposite, while there is no spin-density along other directions, therefore, we have an altermagnet with zero magnetization in the relativistic limit. We turn the N\'eel vector along the $x$-axis and we obtain the spin-density reported in Fig. \ref{fig:spindensity}(b). The spins S$_x$ of Cr$_\uparrow$ and Cr$_\downarrow$ are not equal and opposite, therefore, we have two different spins while the other spin components are zero. In Fig. \ref{fig:spindensity}(c), we plot the average and the difference of the spin-densities of Cr$_\uparrow$ and Cr$_\downarrow$ such that the difference is the part related to the weak ferrimagnetism.
While most of the altermagnets host only weak ferromagnetism, this altermagnet can host weak ferrimagnetism when the N\'eel vector is along the $x$-axis while it has weak ferromagnetism with N\'eel vector along the [110]-direction \cite{antonenko2024mirrorchernbandsweyl}. The weak ferrimagnetism is around one order of magnitude smaller than the average spin density of states, therefore, this is larger than previously investigated weak ferromagnets like RuO$_2$, wurtzite MnSe \cite{autieri2023dzyaloshinskii} and MnTe \cite{Ye25}. Despite the light element, this large relativistic effect arises from the crystal symmetries.


In summary, we investigated the structural, electronic, magnetic and topological properties of the MBene Cr$_2$BAl monolayer. We found that Cr$_2$BAl monolayer exhibits an altermagnetic ground state over a wide range of lattice parameter values. 
In this configuration, Cr$_2$BAl is a d$_{x^2-y^2}$ altermagnet, showing a non-relativistic spin-splitting in the band structure, with opposite splitting along ${\Gamma}$-X and ${\Gamma}$-Y and with degenerate bands along ${\Gamma}$-M.
When the SOC interaction is included and when the N\'eel vector is along $x$, the system hosts a large weak ferrimagnetism if compared with other investigated altermagnets rising from the crystal symmetries, despite the light element.
While, without SOC, the band structure presents multiple crossings and anti-crossings along ${\Gamma}$-X and ${\Gamma}$-Y, the inclusion of SOC opens band gaps around these crossings. 
The compound is an altermagnetic topological crystalline insulator, with the splitting that produces a peak in the spin Hall conductivity and generates Dirac dispersions visible along the [100] surface. 
Building on our findings, the Cr$_2$BAl monolayer emerges as a promising platform for next-generation spintronic and quantum devices, particularly due to its altermagnetic properties and topological features. Given its non-relativistic spin splitting and zero net magnetization, it could be effectively utilized in MTJs to enhance the TMR effect, as it was already shown in other Cr-based MBenes \cite{Sun25}.
Our investigations with SOC interactions further support the fact that Cr$_2$BAl can be explored experimentally for spintronic applications; however, the importance of surface terminations is essential to realize efficient performance in experimental scenarios.\\

\begin{acknowledgments}
We acknowledge W. Sun for the useful discussions. This research was supported by the "MagTop" project (FENG.02.01-IP.05-0028/23) carried out within the "International Research Agendas" programme of the Foundation for Polish Science, co-financed by the European Union under the European Funds for Smart Economy 2021-2027 (FENG).
R.M.S. acknowledges MUR-PRIN2022-CRESO Project for funding. C.A. and G.C. acknowledge support from PNRR MUR project PE0000023-NQSTI. We further acknowledge access to the computing facilities of the Interdisciplinary Center of Modeling at the University of Warsaw, Grant g91-1418, g91-1419, g96-1808 and g96-1809 for the availability of high performance computing resources and support. We acknowledge the access to the computing facilities of the Poznan Supercomputing and Networking Center, Grants No. pl0267-01, pl0365-01 and pl0471-01. R.M.S. acknowledges the CINECA award under the ISCRA initiative IsCc4 "CHARPH2", IsCb7 "CHARPHEN", IsCc1 "DRIFT" and IsB29 "DASP" grants.
\end{acknowledgments}

\bibliography{biblio}
\end{document}